\documentclass[aps,showpacs]{revtex4}
\usepackage{amsmath}
\usepackage{graphicx}
\begin{document}


\title{Neutrino-$^{12}$C scattering in the {\it ab initio} shell model 
with a realistic three-body interaction}

\author{A. C. Hayes$^1$, P. Navr\'atil$^2$ and J. P. Vary$^3$}
\affiliation{$^1$Theoretical Division, Los Alamos National Laboratory, 
Los Alamos, New Mexico 87545 \\
$^2$Lawrence Livermore National Laboratory, L-414, P.O. Box 808, 
Livermore, CA  94551 \\
$^3$Department of Physics and Astronomy, Iowa State University, 
Ames, IA 50011}

\begin{abstract}
We investigate cross sections for neutrino-$^{12}$C exclusive scattering 
and for muon capture 
on $^{12}$C using wave functions obtained in the {\it ab initio} 
no-core shell model. In our parameter-free calculations with basis
spaces up to the $6\hbar\Omega$ we show
that realistic nucleon-nucleon interactions, like e.g. the CD-Bonn,
under predict the experimental cross sections by more than a factor of two.
By including a realistic three-body interaction, Tucson-Melbourne TM$^\prime$(99),
the cross sections are enhanced significantly
and a much better agreement with experiment is achieved. At the same time,
the TM$^\prime$(99) interaction improves the calculated level ordering 
in $^{12}$C. The comparison between the CD-Bonn and the three-body calculations
provides strong confirmation for the need to include a realistic
three-body interaction to account for the spin-orbit strength in 
$p$-shell nuclei.       
\end{abstract}
\maketitle

The Gamow-Teller (GT) transition from the ground state of $^{12}$C to the 1$^+$ T=1 isobar triplet
($^{12}$B$_{g.s.}$, $^{12}$C(15.11 MeV), $^{12}$N$_{g.s.}$) is a very 
sensitive test of nuclear structure models for mass 12
 and, particularly, of the strength of the
spin-orbit interaction.
The two most common $p$-shell approximations for the structure of the ground state of
$^{12}$C ((a) the $p$-shell equivalent of a L=0 S=0 three
alpha-cluster structure and (b) the closed $p_{3/2}$ shell structure) 
give very different (indeed opposite) predictions for the B(GT) strength to T=1 1$^+$ triplet. 
In the $p$-shell alpha-cluster limit the ground state of 
carbon has good SU(4) symmetry [444] 
and the Gamow-Teller
transition is forbidden because there does not exist a 1$^+$ T=1 state with [444] symmetry and
$\sigma\tau$ operator cannot change SU(4) symmetry. This translates into an exact cancellation
between the different $p_{1/2}$ and $p_{3/2}$ transition amplitudes. The observed transition strength
requires the inclusion of higher
SU(4)
components in the wave functions and  
the breaking of the cancellation is quite sensitive to the assumed spin-orbit interaction.
In the the $jj$-coupling limit, where one assumes that the ground state of $^{12}$C is
described by a closed $p_{3/2}$ shell, 
the transition to the T=1 1$^+$ state is pure $p_{3/2}\rightarrow p_{1/2}$. No cancellations between
different transition amplitudes are allowed and the transition
strength is over estimated by almost a factor of 6. When RPA correlations are
included in the initial and final states the situation improves somewhat, 
but the transition remains over-estimated
by about a factor of 4 \cite{kolbe,nu_C12_SM}.
The strong contrast between the predictions of the pure $jj$-coupling and the pure SU(4) limits
makes this Gamow-Teller transition an ideal test case for the strength of the spin-orbit
interaction and for model wave functions of mass 12.

In this letter we present the predictions of no-core shell model (NCSM) \cite{C12_NCSM} calculations 
of $^{12}$C for the T=1 1$^+$ transition in $^{12}$C. We examine  
inelastic electron scattering to the 15.11 MeV state of $^{12}$C, muon capture to the 
ground state of $^{12}$B, and neutrino scattering to the ground  state of $^{12}$N.
These different electroweak reactions probe different momentum transfers
and comparisons between theory and experiment allow us to test the convergence 
of the no-core shell model with increasing basis size, up to 6$\hbar\Omega$. 
We also investigate the contributions of a three-nucleon force since it 
is now well established \cite{pieper,carlson,wiringa} that 
realistic nucleon-nucleon interactions alone
account for only half the observed 
$p$-shell splitting, while the rest arises form two-pion exchange between three or more
nucleons.
In the present calculations we include a realistic chiral-symmetry-based 
three-nucleon interaction (TNI),
Tucson-Melbourne TM$^\prime$(99) \cite{TMprime99}.

A detailed description of the NCSM approach was presented, e.g. in Refs. \cite{C12_NCSM,v3eff}.
Here, we simply present extensions and modifications needed when a genuine
TNI is included. 
The starting Hamiltonian is
%
$H_A= 
\frac{1}{A}\sum_{i<j}\frac{(\vec{p}_i-\vec{p}_j)^2}{2m}
+ \sum_{i<j}^A V_{{\rm NN}, ij} + \sum_{i<j<k}^A V_{{\rm NNN}, ijk}$, 
%
where 
$V_{{\rm NN}, ij}$ is
the nucleon-nucleon (NN) interaction 
and $V_{{\rm NNN}, ijk}$ is the TNI. We employ a large
but finite harmonic-oscillator (HO) basis. Due to properties of the realistic nuclear 
interaction 
we have to derive an effective interaction appropriate for the selected finite basis space.
To facilitate this, we modify the Hamiltonian 
by adding to it the center-of-mass (CM) HO Hamiltonian
$H_{\rm CM}=T_{\rm CM}+ U_{\rm CM}$, where
$U_{\rm CM}=\frac{1}{2}Am\Omega^2 \vec{R}^2$,
$\vec{R}=\frac{1}{A}\sum_{i=1}^{A}\vec{r}_i$.
The effect of the HO CM Hamiltonian will later be subtracted
in the final many-body calculation. 
The modified Hamiltonian can be cast into the form
%
$H_A^\Omega = H_A + H_{\rm CM}=\sum_{i=1}^A h_i + \sum_{i<j}^A V_{ij}^{\Omega,A}
+\sum_{i<j<k}^A V_{{\rm NNN}, ijk}$,
%
where $h_i= \frac{\vec{p}_i^2}{2m}+\frac{1}{2}m\Omega^2 \vec{r}^2_i$
and $V_{ij}^{\Omega,A}=V_{{\rm NN}, ij}-\frac{m\Omega^2}{2A}(\vec{r}_i-\vec{r}_j)^2$.
Next we divide the $A$-nucleon infinite HO basis space
into the finite active space ($P$) comprising all states up to $N_{\rm max}$
HO excitations above the unperturbed ground state and the excluded spaces ($Q=1-P$).  
The basic idea of the NCSM approach is to apply a unitary transformation
on the modified Hamiltonian, 
$e^{-S} H_A^\Omega e^S$ such that
$Q e^{-S} H_A^\Omega e^S P=0$. If such a transformation is found, the effective
Hamiltonian that exactly reproduces a subset of eigenstates of the full space Hamiltonian
is given by $H_{\rm eff}=P e^{-S} H_A^\Omega e^S P$. This effective Hamiltonian
contains up to $A$-body terms and it is essentially as difficult to construct it as to solve
the full problem. Therefore, we apply this approach with a cluster approximation.
When a genuine TNI is considered, the simplest cluster approximation produces 
a three-body effective interaction. The NCSM calculation is then performed in four steps:

(i) We solve a three-nucleon system for all possible three-nucleon channels
with the Hamiltonian $H_A^\Omega$,
i.e.,
using $h_1+h_2+h_3+V_{12}^{\Omega,A}+V_{13}^{\Omega,A}+V_{23}^{\Omega,A}+V_{{\rm NNN}, 123}$.
Consequently, the three nucleons
feel a pseudo-mean field of the spectator nucleons generated by the HO CM potential.
It is necessary to separate the three-body effective interaction 
contributions from the TNI and from the two-nucleon interaction. Therefore, we need to find
three-nucleon solutions for the Hamiltonian with and without the $V_{{\rm NNN}, 123}$ TNI term.
The three-nucleon solutions are obtained by procedures described in Refs. 
\cite{Jacobi_NCSM} (without TNI) and \cite{NCSM_TM} (with TNI).

(ii) We construct the unitary transformation 
corresponding to the choice of the active basis space $P$
from the three-nucleon solutions using the Lee-Suzuki 
procedure \cite{LS1,UMOA}. The three-body effective interaction is then obtained
as $V_{{\rm 3eff},123}^{\rm NN+NNN}=P [e^{-S_{\rm NN+NNN}}(h_1+h_2+h_3+V_{12}^{\Omega,A}
+V_{13}^{\Omega,A}+V_{23}^{\Omega,A}+V_{{\rm NNN}, 123})e^{S_{\rm NN+NNN}}
-(h_1+h_2+h_3)] P$ and
$V_{{\rm 3eff},123}^{\rm NN}=P [e^{-S_{\rm NN}}(h_1+h_2+h_3+V_{12}^{\Omega,A}
+V_{13}^{\Omega,A}+V_{23}^{\Omega,A})e^{S_{\rm NN}}
-(h_1+h_2+h_3)] P$. The three-body effective interaction contribution
from the TNI is then defined as 
$V_{{\rm 3eff},123}^{\rm NNN}\equiv V_{{\rm 3eff},123}^{\rm NN+NNN}-V_{{\rm 3eff},123}^{\rm NN}$.

(iii) As the three-body effective interactions are derived in the Jacobi-coordinate
HO basis but the $A=12$ calculations will be performed in a Cartesian-coordinate
single-particle Slater-determinant m-scheme basis, we need to perform a suitable 
transformation of the interactions. This transformation is a generalization
of the well-known transformation on the two-body level that depends on HO 
Brody-Moshinsky brackets. 

(iv) We solve the Schr\"odinger equation for the $A=12$ nucleon system
using the Hamiltonian 
$H^\Omega_{A, {\rm eff}}=\sum_{i=1}^A h_i + \frac{1}{A-2}\sum_{i<j<k}^A V_{{\rm 3eff}, ijk}^{\rm NN}
+\sum_{i<j<k}^A V_{{\rm 3eff}, ijk}^{\rm NNN}$, where the $\frac{1}{A-2}$ factor takes
care of overcounting the contribution from the two-nucleon interaction. At this point
we also subtract the $H_{\rm CM}$.
The $A=12$ nucleon calculation is then
performed using the Many-Fermion Dynamics shell model code \cite{MFD} generalized to handle three-body
interactions. Eventually, the transition densities are computed that serve as an input for
evaluating our selected observables.

Detailed $^{12}$C NCSM calculations using realistic two-nucleon interactions
were reported in Ref. \cite{C12_NCSM}. Here we extend 
those calculations by including the TNI and reach the $4\hbar\Omega$ ($6\hbar\Omega$)
basis in calculations with (without) the TNI. In Table \ref{tab_c12} we summarize
some of our results. In general, 
in addition to increase of binding energy, we observe a substantial sensitivity 
of the low-lying spectra to the presence of the TNI and 
a trend toward level-ordering and level-spacing improvement in comparison to 
experiment. The sensitivity is the largest for states where the spin-orbit 
interaction strength is known to play a role. Note the correct ordering of the 
$1^+ 0 \leftrightarrow 4^+ 0$ states and ordering and spacing improvement 
of the lowest T=1 states.

The significant increase in the spin-orbit splitting obtained from the
inclusion of the TNI is seen most strikingly
in the predicted cross sections to the T=1 1$^+$ states.
In all our electron scattering and weak interactions results here, only  
one-body currents are included and the bare operators are used. 
Dubach and Haxton \cite{dubach} have shown 
that at high-momentum-transfers it is necessary to include 
two-body meson-exchange currents to describe the transverse magnetic 
electron scattering
form factor for excitation of the 15.11 MeV state. 
However, a reasonable description of the form factor up to
momentum transfers of about 200 MeV/c can be obtained with
a one-body current.

\begin{figure}
\vspace*{2cm}
\includegraphics{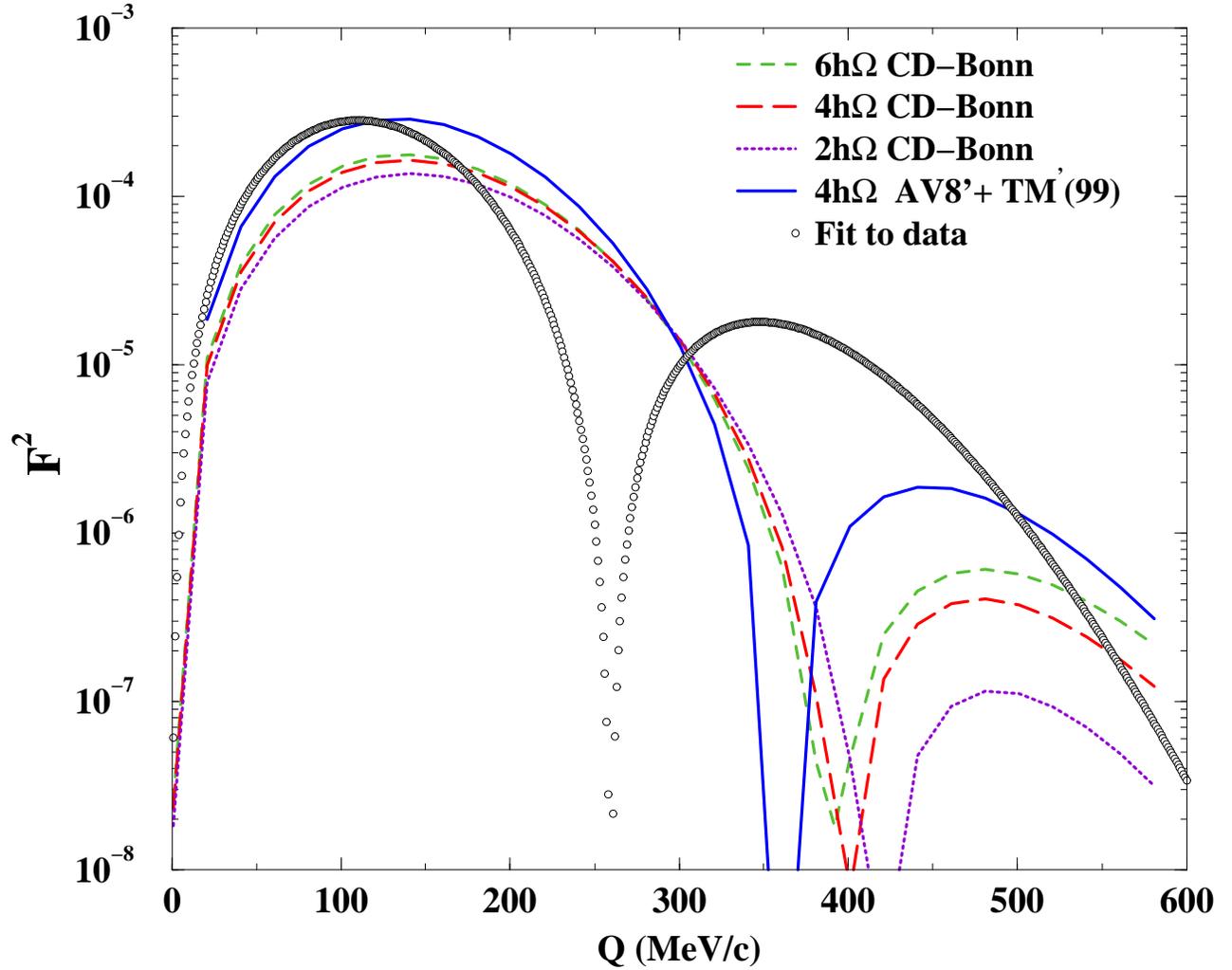}
\caption{\label{formf} The transverse magnetic electron scattering form factor
for the 15.11 MeV T=1 $1^+$ state in $^{12}$C.}
\end{figure}

Figure \ref{formf} shows a comparison between the form factors predicted by
the NCSM and experiment. 
The experimental data are represented by the black circles, which 
represent a fit to the data assuming only a one-body current 
obtained by Dubach and Haxton \cite{dubach}.
The theoretical curves shown are NCSM results for
$2\hbar\Omega$, $4\hbar\Omega$, $6\hbar\Omega$ using the CD-Bonn NN
interaction and a $4\hbar\Omega$ calculation using the 
AV8$^\prime$ plus the TM$^\prime$(99) realistic TNI.
The qualitative features of our results
are seen by looking at the height of the first maximum and
the position of the minimum.
With two-body interactions alone, the change in transition form factor from
$4\hbar\Omega$ to $6\hbar\Omega$ is small compared to the differences
between theory and experiment. The
magnitude of the form factor is too low and minimum occurs too far out in momentum.
When the TNI is included a significant improvement 
is seen in both the shape and magnitude of the theoretical form factor.
The magnitude of the form factor up to the first maximum is close but somewhat lower than
experiment.  The shape of the form factor is also improved
but it is still stretched out too far in momentum.
Comparing the $4\hbar\Omega$ 3-body calculation with the $4\hbar\Omega$ 
CD-Bonn calculation the magnitude of the form factor at the peak 
has increased by about 75$\%$, and the position of the first minimum has shifted
from $q_{\rm min} \sim$ 400 MeV/c to 360 MeV/c. This improvement
is almost entirely due to the improved strength
of the spin-orbit splitting when the TNI is included.

The 15.11 MeV state was included in the fit to the $p$-shell
interaction by Cohen and Kurath \cite{CK} (CK),  
and the CK interaction probably
represents the best description of this transition 
using a globally fitted $p$-shell interaction. 
Our AV8$^\prime$+TM$^\prime$(99) form factor agrees well with the CK prediction, when
for consistency between the two calculations, we use b=1.663 fm. 

\begin{figure}
\includegraphics[width=8.0in]{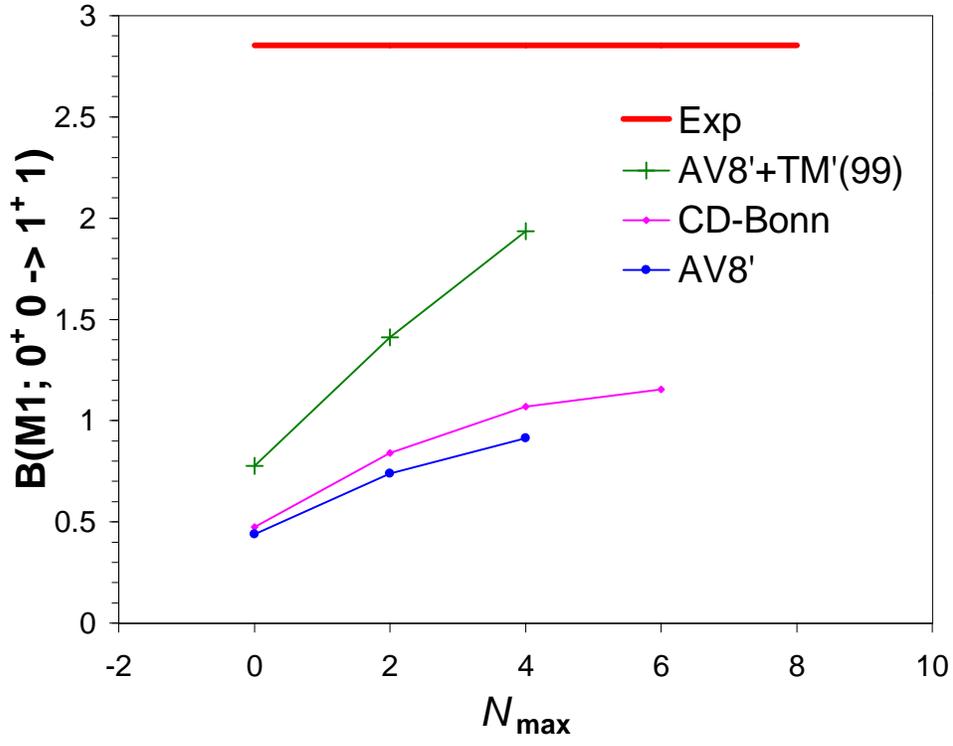}
\caption{\label{c12_bm1} B(M1) values, in $\mu_ N^2$, of the $0^+ 0\rightarrow 1^+ 1$ 
transition in $^{12}$C. For details see Table \protect\ref{tab_c12}.
}
\end{figure}

The conclusion drawn from the transverse magnetic form factor results is further supported
by our B(M1; $1^+ 1\rightarrow 0^+ 0$) results presented in Table \ref{tab_c12} and Fig. \ref{c12_bm1}.
The calculations with 2-body forces show saturation and underpredict the experiment 
by almost a factor of three. By including the TNI, the B(M1) value increases dramatically.
We fully expect that further increases in the basis size will produce
results with TNI close to experiment.  
For smaller basis sizes,
effective transition operators may be important and work in this direction
is underway.
 
Table \ref{rates} shows the comparison between the theoretical and
experimental neutrino scattering cross section for the same selections of
Hamiltonians and basis spaces.  These results show a similar trend to the
electron scattering results above. In this case, 
the neutrino spectrum for electron neutrinos from decay-at-rest (DAR) of the pion peaks around
30 MeV and the average momentum transfer is about 40 MeV/c.
The CD-Bonn interaction 
(without TNI) results indicate an approach to convergence
by $6\hbar\Omega$ but experiment is under-predicted by about a factor of 2.4.
When the TNI is included with the AV8$^\prime$ interaction
the predicted cross section is only 30$\%$ lower than experiment. 
Based on the similarity of trends with the electron scattering results, 
we anticipate that when the model space is eventually
expanded to $6\hbar\Omega$ theory would be within 15$\%$ of experiment.
The substitution of AV8$^\prime$ for CD-Bonn in the calculations 
with TNI is expected to be of minor consequence.

Muon capture involves a higher momentum transfer than the $(\nu_e,e^-)$ reaction and the average momentum
transfer is  $q\sim$ 100 MeV/c. 
By $6\hbar\Omega$ the CD-Bonn calculations show signs of converging yet experiment is underestimated a factor
of  2.6. The inclusion of the TNI shows a significant improvement and, for $4\hbar\Omega$,
theory is 34$\%$ lower than experiment. Again extrapolating using the trends of the
inelastic electron scattering results
suggests that a 6$\hbar\Omega$ calculation that included a realistic TNI would come
within 20$\%$ of experiment.

The $(\nu_\mu,\mu^-)$ neutrino cross section to $^{12}$N$_{g.s.}$  corresponds to the LSND muon neutrinos
from decay-in-flight (DIF) of the pion. This spectrum involves neutrinos up to about 250 MeV, with a average
neutrino energy of about 150 MeV and an average momentum transfer of about 200 MeV/c.
In this case the $6\hbar\Omega$ CD-Bonn calculation is off by a factor of 1.8 compared with experiment.
The $4\hbar\Omega$ calculation that includes the 3-body TM$^\prime$(99)  interaction is, in fact, in agreement
with experiment. However, based on the trends established above, this suggests 
that a larger model space may over-predict experiment. Examining the elastic 
scattering form factor suggests that the problem lies in the fact that 
at 200 MeV/c the predicted form factor is too large. Of course, as the model space 
is increased we expect the form factor to be shifted down in momentum.

In conclusion,
the transition from the $^{12}$C$_{g.s.}$ to the T=1 1$^+$ states in mass 12 is
very sensitive to the strength of the spin orbit interaction.
We have investigated neutrino-$^{12}$C exclusive cross sections and muon capture
on $^{12}$C  as well as inelastic electron scattering 
using wave functions obtained in the {\it ab initio}
NCSM. In our parameter-free calculations with basis
spaces up to $6\hbar\Omega$ we show that realistic NN interactions
under predict the experimental weak interaction  cross sections by more than a factor of two.
At high momentum transfers around $q\sim$ 200 MeV/c the electron scattering form factor is 
over-predicted and the position of the predicted minimum to close to 400 MeV/c, compared
to the experimental minimum at $q\sim 260$ MeV/c.
By including a realistic TNI 
the weak interaction cross sections are enhanced significantly
which considerably improves agreement with experiment. The shape of the electron scattering 
form factor is also significantly improved, but the predicted form factor still peaks  at too
large a momentum transfer and is too large at the momentum transfers relevant to the LSND DIF cross section.
The difference between the observed and predicted shape for
the (e,e') form factors and the very different momentum transfers involved in the
three weak processes examined here
imply that a single experiment/theory scale factor
cannot be defined for all three. 
The comparison between the CD-Bonn and the three-body calculations discussed here
provide a strong
confirmation of the need to include a realistic three-body interaction to account for the 
spin-orbit strength in $p$-shell nuclei.

This work was performed in part under the auspices of
the U. S. Department of Energy by the University of California,
Lawrence Livermore National Laboratory under contract
No. W-7405-Eng-48. This work was also supported in part by USDOE
DE-FG02-87ER40371.

\begin{table}
\begin{tabular}{c|c|c|c|cccc}
 & $^{12}$C & AV8$^\prime$+TM$^\prime$(99) & AV8$^\prime$  & & CD-Bonn &  &    \\ 
 basis space & - &$4\hbar\Omega$ &$4\hbar\Omega$ & $6\hbar\Omega$ & $4\hbar\Omega$ & $2\hbar\Omega$ 
& $0\hbar\Omega$ \\
\hline
$|E_{\rm gs}|$ [MeV]   & 92.162 & 91.963 & 85.944 & 85.630 & 88.518 & 92.375 & 104.947  \\
$Q_{2^+}$ [$e$ fm$^2$] & +6(3)  & 4.288 & 4.613 & 4.717  & 4.532  & 4.430  & 4.253    \\
\hline
$E_{\rm x}(2^+ 0)$ [MeV] & 4.439  & 3.603  & 3.427  & 3.612  & 3.697  & 3.837   & 3.734  \\
$E_{\rm x}(1^+ 0)$ [MeV] & 12.710 & 11.280 & 13.926 & 13.930 & 14.140 & 14.524  & 13.866 \\
$E_{\rm x}(4^+ 0)$ [MeV] & 14.083 & 13.517 & 12.272 & 13.110 & 13.356 & 13.638  & 12.406 \\
$E_{\rm x}(1^+ 1)$ [MeV] & 15.110 & 16.221 & 16.364 & 16.064 & 16.165 & 16.291  & 15.290 \\
$E_{\rm x}(2^+ 1)$ [MeV] & 16.106 & 16.467 & 17.712 & 17.409 & 17.717 & 17.945  & 15.970 \\
$E_{\rm x}(0^+ 1)$ [MeV] & 17.760 & 17.116 & 16.213 & 16.534 & 16.619 & 16.493  & 14.698 \\
\hline
B(E2;$2^+0 \rightarrow 0^+0$) & 7.59(42)  & 4.146  & 4.765  & 5.019  & 4.624  & 4.412  & 4.092  \\
B(M1;$1^+1 \rightarrow 0^+0$) & 0.951(20) & 0.645  & 0.305  & 0.384  & 0.355  & 0.280  & 0.158  \\
B(E2;$2^+1 \rightarrow 0^+0$) & 0.65(13)  & 0.430  & 0.247  & 0.309  & 0.283  & 0.015  & 0.002 \\
\end{tabular}
\caption{\label{tab_c12} Experimental and calculated properties
of $^{12}$C. The units are $e^2$ fm$^4$ ($\mu_N^2$) for B(E2) (B(M1)). 
Three-body effective interactions derived from
the AV8$^\prime$ \protect\cite{av8p} and AV8$^\prime$+TM$^\prime$(99) 
and two-body effective interactions
derived from the CD-Bonn \protect\cite{cdb} NN potential and a 
HO frequency of $\hbar\Omega=15$ MeV were used.
The TM$^\prime$(99) parameters are given in Refs. \protect\cite{TMprime99,NCSM_TM}
with the cutoff set to $\Lambda=4.7$. 
The experimental values are from Ref. \protect\cite{AS90}.
By extrapolating our results we predict the CD-Bonn $^{12}$C binding energy to be
$\approx 80\pm 2$ MeV.
}
\end{table}

\begin{table}
\begin{tabular}{c|ccccc}
Interaction   & CD-Bonn 2$\hbar\Omega$ & CD-Bonn 4$\hbar\Omega$&
 CD-Bonn 6$\hbar\Omega$ &AV8$^\prime$+TM$^\prime$(99) $4\hbar\Omega$ & experiment \\
\hline
 $(\nu_e,e^-)$ & 2.27&    3.2&    3.69&     6.8  & 8.9$\pm$0.3$\pm$0.9 \protect\cite{nu_exp_DAR} \\
 $(\nu_\mu,\mu^-)$&0.168 &  0.275&  0.312 &   0.537  &    0.56$\pm$0.08$\pm$0.1 \protect\cite{nu_exp_DIF} \\
$\mu$-capture& 1.46 &   2.07&   2.38 &    4.43  &    6.0$\pm$0.4 \protect\cite{mu_exp} \\  
\end{tabular}
\caption{\label{rates} Predicted weak interaction rates for 
the $^{12}$C$\rightarrow$ T=1 $1^+$ transitions. 
The units are $10^{-42} {\rm cm}^2$ for the $(\nu_e,e^-)$ DAR  
cross section, $10^{-40} {\rm cm}^2$ for $(\nu_\mu,\mu^-)$ DIF cross 
section and $10^3{\rm sec}^{-1}$ for muon capture.}
\end{table}

\end{document}